\documentclass{aastex631}

\def\plus{\texttt{+}}
\def\minus{\texttt{-}}
\usepackage{mathastext}
\usepackage{textcomp}
\usepackage{pgfplots}
\pgfplotsset{compat=1.9}

\shortauthors{Kiwy et al. 2021}

\begin{document}

\title{Discovery of a low-mass comoving system using NOIRLab Source Catalog DR2}

\author[0000-0001-8662-1622]{Frank Kiwy}
\affiliation{Backyard Worlds: Planet 9}
\author[0000-0001-6251-0573]{Jacqueline Faherty}
\affiliation{Department of Astrophysics, American Museum of Natural History, Central Park West at 79th Street, NY 10024, USA}
\author[0000-0002-1125-7384]{Aaron Meisner}
\affiliation{NSF's National Optical-Infrared Astronomy Research Laboratory, 950 N. Cherry Ave., Tucson, AZ 85719, USA}
\author[0000-0002-6294-5937]{Adam C. Schneider}
\affiliation{United States Naval Observatory, Flagstaff Station, 10391 West Naval Observatory Rd., Flagstaff, AZ 86005, USA}
\affiliation{Department of Physics and Astronomy, George Mason University, MS3F3, 4400 University Drive, Fairfax, VA 22030, USA}
\author[0000-0002-2387-5489]{Marc Kuchner}
\affiliation{NASA Goddard Space Flight Center, Exoplanets and Stellar Astrophysics Laboratory, Code 667, Greenbelt, MD 20771, USA}
\author[0000-0003-4269-260X]{J. Davy Kirkpatrick}
\affiliation{IPAC, Mail Code 100-22, Caltech, 1200 E. California Blvd., Pasadena, CA 91125, USA}
\author{The Backyard Worlds: Planet 9 Collaboration}

\begin{abstract}
We present the discovery of a low-mass comoving system found by means of the NOIRLab Source Catalog (NSC) DR2. The system consists of the high proper-motion star LEHPM 5005 and an ultracool companion 2MASS J22410186-4500298 with an estimated spectral type of L2. The primary (LEHPM 5005) is likely a mid-M dwarf but over-luminous for its color, indicating a possible close equal mass binary. According to the {\it Gaia} EDR3 parallax of the primary, the system is located at a distance of $58\pm2$ pc. We calculated an angular separation of $7\farcs2$ between both components, resulting in a projected physical separation of 418 AU. 
\end{abstract}

\keywords{Low mass stars(2050) --- Binary stars(154)}

\section*{Introduction}
The NSC \citep{2021AJ....161..192N} is a catalog of most of the public image data in NOIRLab's Astro Data Archive. These images come from telescopes in both hemispheres (CTIO\minus 4m\plus DECam, KPNO\minus 4m\plus Mosaic3 and Bok\minus 2.3\plus 90Prime) and cover $\sim35,000$ square degrees of the sky. The second NSC public data release includes more than 3.9 billion single objects with over 68 billion individual source measurements. It has depths of $\sim23rd$ magnitude in most broadband filters, accurate proper motions and an astrometric accuracy of $\sim2$ mas. The NSC can measure motions to several magnitudes beyond the {\it Gaia} limit for sources of all colors, which makes it an excellent resource for discovering rare moving late-type objects and their possible companions.

\section*{Discovery method}
The system has been discovered by cross-matching a list of previously identified brown dwarf candidates with the NSC DR2 catalog, and by comparing their respective proper motions. The list of brown dwarf candidates was obtained by employing NSC DR2 proper motions and photometry. Using the z, i and Y filter bandpasses along with the relations described in \cite{2019MNRAS.489.5301C}, we were able to determine an appropriate color cut for the brown dwarf selection. Candidates having proper motions below 100 mas yr$^{-1}$ have been discarded from the selection to reduce the number of false positives and to provide proper motions significant enough to be visually confirmed by blinking images of different epochs.

By following this method, we identified the high-proper motion star known as LEHPM 5005 and its ultracool companion 2MASS J2241-4500.

\section*{System characteristics}
Both components of the system have NSC DR2 detections with high significance proper motions. We determined a difference in proper motion between the primary and the secondary of less than 0.5 mas yr$^{-1}$ in RA and less than 3.8 mas yr$^{-1}$ in DEC. Furthermore, we find that the primary's proper motions from {\it Gaia} EDR3 \citep{2021AA...649A...1G} seem to confirm the NSC DR2 measurements (Table \ref{tab:param} \hyperlink{tab_param_astro} {Astrometry}).

We calculated a distance of $58\pm2$ pc for the primary using its {\it Gaia} EDR3 parallax. Although the secondary has {\it Gaia} DR2 and EDR3 detections, neither data release includes a parallax measurements for this source. Since the secondary is listed in 2MASS and VISTA VHS DR6 \citep{2021yCat.2367....0M}, we can use the $J-K_s$ color to perform a spectral type estimate and calculate its photometric distance. We determined a spectral type of L3 for the 2MASS $J-K_s$ color and a spectral type of L2 for the VHS DR6 $J-K_s$ color, resulting in photometric distances between 47.8 and 51.3 pc using the 2MASS photometry and between 59.8 and 60.2 pc using the VHS DR6 photometry. The spectral type estimates and corresponding absolute magnitudes, necessary to calculate the photometric distances, were derived from the relations reported in \cite{2018ApJS..234....1B}.

We used the NSC DR2 positions of both components to compute their angular separation of $7\farcs2$, corresponding to a projected physical separation of 418 AU. The position of the primary was corrected for proper motion to the epoch of the secondary. Using the {\it Gaia} EDR3 parallax and proper motions of the primary, we calculated a tangential velocity of $87\pm3$ km s$^{-1}$ for this system.

We estimated spectral types between M4 and M6 for the primary and between L0 and L4 for the secondary, resulting in a mean spectral type of M5.5 for the primary and L2 for the secondary (Table \ref{tab:param} \hyperlink{tab_param_spt} {Spectral types}). These spectral type estimates were obtained by employing {\it Gaia} EDR3, NSC DR2, 2MASS, VISTA VHS DR6 and AllWISE photometry (Table \ref{tab:param} \hyperlink{tab_param_phot} {Photometry}) and by applying the relations described in \cite{2019AJ....157..231K}, \cite{2019MNRAS.489.5301C} and \cite{2018ApJS..234....1B}.

Physical properties of the primary were determined in \cite{2019AJ....158..138S} and given in Table \ref{tab:param} \hyperlink{tab_param_fund} {Fundamentals}.

\section*{Chance alignment probability}
We calculated two chance alignment probabilities, one without a distance constraint and one with the previously calculated photometric distances of the secondary.

If we use no distance constraint and a proper motion matching tolerance of $\pm10$ mas yr$^{-1}$, we find that the secondary has 52 {\it Gaia} EDR3 proper motion matches over the entire sky, corresponding to a chance alignment probability of $3.61\times10^{-5}$ that one of these 52 would randomly happen to fall within a radius corresponding to the distance at which a mid-M dwarf could retain, with $\sim50\%$ probability, a companion over the age of the system. For this purpose, we assume an age of $\sim5$ Gyr, which for a solar-type star corresponds to a separation of about 0.1 pc ($\sim20,000$ AU).

If we additionally put in a distance constraint of 47.8 pc $<$ d $<$ 60.2 pc, this decreases the number of {\it Gaia} EDR3 matches to 9 over the entire sky and the chance alignment probability accordingly goes down to $6.25\times10^{-6}$.

\vspace{3mm}
Alternatively, we used the CoMover code from \cite{2021ascl.soft06007G}, which is based on models instead of catalog data, and obtained a chance alignment probability of $6.65\times10^{-6}$. This result appears to be consistent with the chance alignment probability previously calculated using a distance constraint.

\section*{Discussion}
We found that the primary has a substantially high {\it Gaia} RUWE \footnote{Re-normalized Unit Weight Error} of 7.9 in DR2 and 14.0 in EDR3. According to the {\it Gaia} DR2 documentation, the RUWE is expected to be around 1.0 for sources where the single-star model provides a good fit to the astrometric observations. A value significantly greater than 1.0 could indicate that either the source is non-single, or problematic for the astrometric solution. Given the system's angular separation of $7\farcs2$ and the {\it Gaia} effective angular resolution of $0\farcs4$ (DR2), we can exclude that the high RUWE is caused by the actual secondary.

The primary is over-luminous for its color, indicating that it is very likely a close binary. Its position on the {\it Gaia} color magnitude diagram and its high RUWE \citep{2020MNRAS.496.1922B} \citep{2020MNRAS.495..321P}, increasing between data releases, seem to confirm that hypothesis.

Furthermore, the primary finds mention in \cite{2020yCat..51590060G}, a catalog of 8695 flares from 1228 stars in TESS sectors 1 \& 2, which may be a further hint of binarity, since close binaries can cause flaring and a more rapid rotation.

\newpage
\section*{Acknowledgements}
This work has made use of data provided by:
\begin{itemize}
  \item The European Space Agency (ESA) mission {\it Gaia} (\url{https://www.cosmos.esa.int/gaia}), processed by the {\it Gaia} Data Processing and Analysis Consortium (DPAC, \url{https://www.cosmos.esa.int/web/gaia/dpac/consortium}). Funding for the DPAC has been provided by national institutions, in particular the institutions participating in the {\it Gaia} Multilateral Agreement.
  \item The Astro Data Lab at NSF's National Optical-Infrared Astronomy Research Laboratory. NOIRLab is operated by the Association of Universities for Research in Astronomy (AURA), Inc. under a cooperative agreement with the National Science Foundation.
\end{itemize}

\begin{deluxetable*}{cccccc}
\tablecaption{Measured Parameters\label{tab:param}}
\tablehead{
\colhead{\footnotesize{\textbf{Parameter}}} &
\colhead{\footnotesize{\textbf{Primary}}} &
\colhead{\footnotesize{\textbf{Secondary}}} &
\colhead{\footnotesize{\textbf{System}}} &
\colhead{\footnotesize{\textbf{Unit}}} &
\colhead{\footnotesize{\textbf{Reference}}} \\
\colhead{} &
\colhead{LEHPM 5005} &
\colhead{2MASS J22410186-4500298} &
\colhead{} &
\colhead{} &
\colhead{}
}
\tabletypesize{\scriptsize}
\decimalcolnumbers
\startdata
\textbf{\hypertarget{tab_param_astro}{ASTROMETRY}} \\
\hline\hline
$\alpha_{NSC}$ & $340.2579184\pm0\farcs002945$ & $340.2595948\pm0\farcs00365$ & \nodata & deg & 1 \\
$\delta_{NSC}$ & $-45.007344\pm0\farcs0032$ & $-45.0089825\pm0\farcs003825$ & \nodata & deg & 1 \\
$\mu_{\alpha_{NSC}}$ & $270.602\pm1.938$ & $271.007\pm1.724$ & \nodata & mas yr$^{-1}$ & 1 \\
$\mu_{\delta_{NSC}}$ & $-150.33\pm2.114$ & $-146.575\pm1.809$ & \nodata & mas yr$^{-1}$ & 1 \\
$\mu_{\alpha_{{\it Gaia}}}$ & $273.035\pm0.404$ & \nodata & \nodata & mas yr$^{-1}$ & 2 \\
$\mu_{\delta_{{\it Gaia}}}$ & $-154.157\pm0.509$ & \nodata & \nodata & mas yr$^{-1}$ & 2 \\
$\varpi_{{\it Gaia}}$ & $17.1738\pm0.6061$ & \nodata & \nodata & mas & 2 \\
\hline\hline
\textbf{\hypertarget{tab_param_phot}{PHOTOMETRY}} \\
\hline\hline
G$_{BP}$ & $17.61\pm0.007$ & \nodata & \nodata & mag & 2 \\
G & $15.668\pm0.004$ & $21.101\pm0.023$ & \nodata & mag & 2 \\
G$_{RP}$ & $14.245\pm0.004$ & $19.38\pm0.089$ & \nodata & mag & 2 \\
g$_{NSC}$ & $18.042\pm0.001$ & \nodata & \nodata & mag & 1 \\
r$_{NSC}$ & $16.452\pm0.001$ & $22.336\pm0.036$ & \nodata & mag & 1 \\
i$_{NSC}$ & \nodata & $20.117\pm0.008$ & \nodata & mag & 1 \\
z$_{NSC}$ & $13.795\pm0.001$ & $18.684\pm0.005$ & \nodata & mag & 1 \\
Y$_{NSC}$ & $13.516\pm0.001$ & $18.154\pm0.014$ & \nodata & mag & 1 \\
J$_{2MASS}$ & $12.237\pm0.024$ & $16.168\pm0.105$ & \nodata & mag & 3 \\
H$_{2MASS}$ & $11.593\pm0.026$ & $15.371\pm0.126$ & \nodata & mag & 3 \\
K$_{s2MASS}$ & $11.269\pm0.023$ & $14.611\pm0.084$ & \nodata & mag & 3 \\
J$_{VHS}$ & $12.246\pm0.001$ & $16.202\pm0.006$ & \nodata & mag & 4 \\
K$_{sVHS}$ & $11.321\pm0.001$ & $14.728\pm0.007$ & \nodata & mag & 4 \\
W1 & $11.103\pm0.025$ & $13.993\pm0.135^a$ & \nodata & mag & 5 \\
W2 & $10.928\pm0.023$ & $13.714\pm0.134^a$ & \nodata & mag & 5 \\
W3 & $10.827\pm0.102$ & $>12.362^a$ & \nodata & mag & 5 \\
W4 & $>8.603$ & $>8.964^a$ & \nodata & mag & 5 \\
\hline\hline
\textbf{\hypertarget{tab_param_spt}{SPECTRAL TYPE ESTIMATES}} \\
\hline\hline
$M_G$ & M5 & \nodata & \nodata & \nodata & 6 \\
$M_{G_{RP}}$ & M5 & \nodata & \nodata & \nodata & 6 \\
$G-G_{RP}$ & M6 & L4 & \nodata & \nodata & 6 \\
$G_{BP}-G_{RP}$ & M6 & \nodata & \nodata & \nodata & 6 \\
$G_{BP}-G$ & M6 & \nodata & \nodata & \nodata & 6 \\
$(i-z)_{NSC}$ & \nodata & L0 & \nodata & \nodata & 7 \\
$(J-K_s)_{2MASS}$ & M6 & L3 & \nodata & \nodata & 8 \\
$(J-K_s)_{VHS}$ & M6 & L1,L2 & \nodata & \nodata & 8 \\
$W1-W2$ & M4 & \nodata & \nodata & \nodata & 8 \\
\hline\hline
\textbf{\hypertarget{tab_param_fund}{FUNDAMENTALS}} \\
\hline\hline
Teff & $3014\pm157$ & \nodata & \nodata & K & 9 \\
log g & $4.9062\pm0.0089$ & \nodata & \nodata & cm s$^{-2}$ & 9 \\
Radius & $0.318\pm0.017$ & \nodata & \nodata & R$_\odot$ & 9 \\
Mass & $0.297\pm0.026$ & \nodata & \nodata & M$_\odot$ & 9 \\
Luminosity & $0.00753\pm0.00241$ & \nodata & \nodata & L$_\odot$ & 9 \\
Density & $9.2360\pm0.6912$ & \nodata & \nodata & $\rho_\odot$ & 9 \\
Rotation period & 4.55 & \nodata & \nodata & d & 10 \\
\hline\hline
\textbf{KINEMATICS} \\
\hline\hline
Distance$^b$ & $58\pm2$ & \nodata & $58\pm2$ & pc & 11 \\
$\upsilon_{tan}${$^c$} & $87\pm3$ & \nodata & $87\pm3$ & km s$^{-1}$ & 11 \\
\hline\hline
\textbf{SYSTEM} \\
\hline\hline
Separation$^d$ & \nodata & \nodata & 7.2 & $\arcsec$ & 11 \\
Projected physical \\ separation& \nodata & \nodata & 418 & AU & 11 \\
\enddata
\hfill \\
$^a$Most likely contaminated by the primary \\
$^b$Calculated using $D = 1 / \pi $ \\
$^c$Calculated using {\it Gaia} EDR3 astrometry \\
$^d$Calculated using NSC DR2 positions, translating the position of the primary to the epoch of the secondary \\
References: (1) \cite{2021AJ....161..192N}, (2) \cite{2021AA...649A...1G}, (3) \cite{2003yCat.2246....0C}, (4) \cite{2021yCat.2367....0M}, (5) \cite{2014yCat.2328....0C}, (6) \cite{2019AJ....157..231K}, (7) \cite{2019MNRAS.489.5301C}, (8) \cite{2018ApJS..234....1B}, (9) \cite{2019AJ....158..138S}, (10) \cite{2020yCat..51590060G}, (11) This work.
\end{deluxetable*}

\bibliography{researchnote}

\begin{thebibliography}{}
\expandafter\ifx\csname natexlab\endcsname\relax\def\natexlab#1{#1}\fi
\providecommand{\url}[1]{\href{#1}{#1}}
\providecommand{\dodoi}[1]{doi:~\href{http://doi.org/#1}{\nolinkurl{#1}}}
\providecommand{\doeprint}[1]{\href{http://ascl.net/#1}{\nolinkurl{http://ascl.net/#1}}}
\providecommand{\doarXiv}[1]{\href{https://arxiv.org/abs/#1}{\nolinkurl{https://arxiv.org/abs/#1}}}

\bibitem[{{Belokurov} {et~al.}(2020){Belokurov}, {Penoyre}, {Oh}, {Iorio},
  {Hodgkin}, {Evans}, {Everall}, {Koposov}, {Tout}, {Izzard}, {Clarke}, \&
  {Brown}}]{2020MNRAS.496.1922B}
{Belokurov}, V., {Penoyre}, Z., {Oh}, S., {et~al.} 2020, \mnras, 496, 1922,
  \dodoi{10.1093/mnras/staa1522}

\bibitem[{{Best} {et~al.}(2018){Best}, {Magnier}, {Liu}, {Aller}, {Zhang},
  {Burgett}, {Chambers}, {Draper}, {Flewelling}, {Kaiser}, {Kudritzki},
  {Metcalfe}, {Tonry}, {Wainscoat}, \& {Waters}}]{2018ApJS..234....1B}
{Best}, W. M.~J., {Magnier}, E.~A., {Liu}, M.~C., {et~al.} 2018, \apjs, 234, 1,
  \dodoi{10.3847/1538-4365/aa9982}

\bibitem[{{Carnero Rosell} {et~al.}(2019){Carnero Rosell}, {Santiago}, {dal
  Ponte}, {Burningham}, {da Costa}, {James}, {Marshall}, {McMahon}, {Bechtol},
  {De Paris}, {Li}, {Pieres}, {Garc{\'\i}a-Bellido}, {Abbott}, {Annis},
  {Avila}, {Bernstein}, {Brooks}, {Burke}, {Carrasco Kind}, {Carretero}, {De
  Vicente}, {Drlica-Wagner}, {Fosalba}, {Frieman}, {Gaztanaga}, {Gruendl},
  {Gschwend}, {Gutierrez}, {Hollowood}, {Maia}, {Menanteau}, {Miquel},
  {Plazas}, {Roodman}, {Sanchez}, {Scarpine}, {Schindler}, {Serrano},
  {Sevilla-Noarbe}, {Smith}, {Sobreira}, {Soares-Santos}, {Suchyta}, {Swanson},
  {Tarle}, {Vikram}, {Walker}, \& {DES Collaboration}}]{2019MNRAS.489.5301C}
{Carnero Rosell}, A., {Santiago}, B., {dal Ponte}, M., {et~al.} 2019, \mnras,
  489, 5301, \dodoi{10.1093/mnras/stz2398}

\bibitem[{{Cutri} {et~al.}(2003){Cutri}, {Skrutskie}, {van Dyk}, {Beichman},
  {Carpenter}, {Chester}, {Cambresy}, {Evans}, {Fowler}, {Gizis}, {Howard},
  {Huchra}, {Jarrett}, {Kopan}, {Kirkpatrick}, {Light}, {Marsh}, {McCallon},
  {Schneider}, {Stiening}, {Sykes}, {Weinberg}, {Wheaton}, {Wheelock}, \&
  {Zacarias}}]{2003yCat.2246....0C}
{Cutri}, R.~M., {Skrutskie}, M.~F., {van Dyk}, S., {et~al.} 2003, VizieR Online
  Data Catalog, II/246

\bibitem[{{Cutri} {et~al.}(2021){Cutri}, {Wright}, {Conrow}, {Fowler},
  {Eisenhardt}, {Grillmair}, {Kirkpatrick}, {Masci}, {McCallon}, {Wheelock},
  {Fajardo-Acosta}, {Yan}, {Benford}, {Harbut}, {Jarrett}, {Lake}, {Leisawitz},
  {Ressler}, {Stanford}, {Tsai}, {Liu}, {Helou}, {Mainzer}, {Gettngs},
  {Gonzalez}, {Hoffman}, {Marsh}, {Padgett}, {Skrutskie}, {Beck}, {Papin}, \&
  {Wittman}}]{2014yCat.2328....0C}
{Cutri}, R.~M., {Wright}, E.~L., {Conrow}, T., {et~al.} 2021, VizieR Online
  Data Catalog, II/328

\bibitem[{{Gagn{\'e}} {et~al.}(2021){Gagn{\'e}}, {Faherty}, {Schneider}, \&
  {Meisner}}]{2021ascl.soft06007G}
{Gagn{\'e}}, J., {Faherty}, J.~K., {Schneider}, A.~C., \& {Meisner}, A.~M.
  2021, {CoMover: Bayesian probability of co-moving stars}.
\newblock \doeprint{2106.007}

\bibitem[{{Gunther} {et~al.}(2020){Gunther}, {Zhan}, {Seager}, {Rimmer},
  {Ranjan}, {Stassun}, {Oelkers}, {Daylan}, {Newton}, {Kristiansen}, {Olah},
  {Gillen}, {Rappaport}, {Ricker}, {Vanderspek}, {Latham}, {Winn}, {Jenkins},
  {Glidden}, {Fausnaugh}, {Levine}, {Dittmann}, {Quinn}, {Krishnamurthy}, \&
  {Ting}}]{2020yCat..51590060G}
{Gunther}, M.~N., {Zhan}, Z., {Seager}, S., {et~al.} 2020, VizieR Online Data
  Catalog, J/AJ/159/60

\bibitem[{{{\it Gaia} Collaboration} {et~al.}(2021){{\it Gaia} Collaboration},
  {Brown}, {Vallenari}, {Prusti}, {de Bruijne}, {Babusiaux}, {Biermann},
  {Creevey}, {Evans}, {Eyer}, {Hutton}, {Jansen}, {Jordi}, {Klioner},
  {Lammers}, {Lindegren}, {Luri}, {Mignard}, {Panem}, {Pourbaix}, {Randich},
  {Sartoretti}, {Soubiran}, {Walton}, {Arenou}, {Bailer-Jones}, {Bastian},
  {Cropper}, {Drimmel}, {Katz}, {Lattanzi}, {van Leeuwen}, {Bakker},
  {Cacciari}, {Casta{\~n}eda}, {De Angeli}, {Ducourant}, {Fabricius},
  {Fouesneau}, {Fr{\'e}mat}, {Guerra}, {Guerrier}, {Guiraud}, {Jean-Antoine
  Piccolo}, {Masana}, {Messineo}, {Mowlavi}, {Nicolas}, {Nienartowicz},
  {Pailler}, {Panuzzo}, {Riclet}, {Roux}, {Seabroke}, {Sordo}, {Tanga},
  {Th{\'e}venin}, {Gracia-Abril}, {Portell}, {Teyssier}, {Altmann}, {Andrae},
  {Bellas-Velidis}, {Benson}, {Berthier}, {Blomme}, {Brugaletta}, {Burgess},
  {Busso}, {Carry}, {Cellino}, {Cheek}, {Clementini}, {Damerdji}, {Davidson},
  {Delchambre}, {Dell'Oro}, {Fern{\'a}ndez-Hern{\'a}ndez}, {Galluccio},
  {Garc{\'\i}a-Lario}, {Garcia-Reinaldos}, {Gonz{\'a}lez-N{\'u}{\~n}ez},
  {Gosset}, {Haigron}, {Halbwachs}, {Hambly}, {Harrison}, {Hatzidimitriou},
  {Heiter}, {Hern{\'a}ndez}, {Hestroffer}, {Hodgkin}, {Holl}, {Jan{\ss}en},
  {Jevardat de Fombelle}, {Jordan}, {Krone-Martins}, {Lanzafame},
  {L{\"o}ffler}, {Lorca}, {Manteiga}, {Marchal}, {Marrese}, {Moitinho}, {Mora},
  {Muinonen}, {Osborne}, {Pancino}, {Pauwels}, {Petit}, {Recio-Blanco},
  {Richards}, {Riello}, {Rimoldini}, {Robin}, {Roegiers}, {Rybizki}, {Sarro},
  {Siopis}, {Smith}, {Sozzetti}, {Ulla}, {Utrilla}, {van Leeuwen}, {van
  Reeven}, {Abbas}, {Abreu Aramburu}, {Accart}, {Aerts}, {Aguado}, {Ajaj},
  {Altavilla}, {{\'A}lvarez}, {{\'A}lvarez Cid-Fuentes}, {Alves}, {Anderson},
  {Anglada Varela}, {Antoja}, {Audard}, {Baines}, {Baker},
  {Balaguer-N{\'u}{\~n}ez}, {Balbinot}, {Balog}, {Barache}, {Barbato},
  {Barros}, {Barstow}, {Bartolom{\'e}}, {Bassilana}, {Bauchet},
  {Baudesson-Stella}, {Becciani}, {Bellazzini}, {Bernet}, {Bertone}, {Bianchi},
  {Blanco-Cuaresma}, {Boch}, {Bombrun}, {Bossini}, {Bouquillon}, {Bragaglia},
  {Bramante}, {Breedt}, {Bressan}, {Brouillet}, {Bucciarelli}, {Burlacu},
  {Busonero}, {Butkevich}, {Buzzi}, {Caffau}, {Cancelliere}, {C{\'a}novas},
  {Cantat-Gaudin}, {Carballo}, {Carlucci}, {Carnerero}, {Carrasco},
  {Casamiquela}, {Castellani}, {Castro-Ginard}, {Castro Sampol}, {Chaoul},
  {Charlot}, {Chemin}, {Chiavassa}, {Cioni}, {Comoretto}, {Cooper}, {Cornez},
  {Cowell}, {Crifo}, {Crosta}, {Crowley}, {Dafonte}, {Dapergolas}, {David},
  {David}, {de Laverny}, {De Luise}, {De March}, {De Ridder}, {de Souza}, {de
  Teodoro}, {de Torres}, {del Peloso}, {del Pozo}, {Delbo}, {Delgado},
  {Delgado}, {Delisle}, {Di Matteo}, {Diakite}, {Diener}, {Distefano},
  {Dolding}, {Eappachen}, {Edvardsson}, {Enke}, {Esquej}, {Fabre}, {Fabrizio},
  {Faigler}, {Fedorets}, {Fernique}, {Fienga}, {Figueras}, {Fouron},
  {Fragkoudi}, {Fraile}, {Franke}, {Gai}, {Garabato}, {Garcia-Gutierrez},
  {Garc{\'\i}a-Torres}, {Garofalo}, {Gavras}, {Gerlach}, {Geyer}, {Giacobbe},
  {Gilmore}, {Girona}, {Giuffrida}, {Gomel}, {Gomez}, {Gonzalez-Santamaria},
  {Gonz{\'a}lez-Vidal}, {Granvik}, {Guti{\'e}rrez-S{\'a}nchez}, {Guy},
  {Hauser}, {Haywood}, {Helmi}, {Hidalgo}, {Hilger}, {H{\l}adczuk}, {Hobbs},
  {Holland}, {Huckle}, {Jasniewicz}, {Jonker}, {Juaristi Campillo}, {Julbe},
  {Karbevska}, {Kervella}, {Khanna}, {Kochoska}, {Kontizas}, {Kordopatis},
  {Korn}, {Kostrzewa-Rutkowska}, {Kruszy{\'n}ska}, {Lambert}, {Lanza}, {Lasne},
  {Le Campion}, {Le Fustec}, {Lebreton}, {Lebzelter}, {Leccia}, {Leclerc},
  {Lecoeur-Taibi}, {Liao}, {Licata}, {Lindstr{\o}m}, {Lister}, {Livanou},
  {Lobel}, {Madrero Pardo}, {Managau}, {Mann}, {Marchant}, {Marconi}, {Marcos
  Santos}, {Marinoni}, {Marocco}, {Marshall}, {Martin Polo},
  {Mart{\'\i}n-Fleitas}, {Masip}, {Massari}, {Mastrobuono-Battisti}, {Mazeh},
  {McMillan}, {Messina}, {Michalik}, {Millar}, {Mints}, {Molina}, {Molinaro},
  {Moln{\'a}r}, {Montegriffo}, {Mor}, {Morbidelli}, {Morel}, {Morris},
  {Mulone}, {Munoz}, {Muraveva}, {Murphy}, {Musella}, {Noval}, {Ord{\'e}novic},
  {Orr{\`u}}, {Osinde}, {Pagani}, {Pagano}, {Palaversa}, {Palicio}, {Panahi},
  {Pawlak}, {Pe{\~n}alosa Esteller}, {Penttil{\"a}}, {Piersimoni}, {Pineau},
  {Plachy}, {Plum}, {Poggio}, {Poretti}, {Poujoulet}, {Pr{\v{s}}a}, {Pulone},
  {Racero}, {Ragaini}, {Rainer}, {Raiteri}, {Rambaux}, {Ramos}, {Ramos-Lerate},
  {Re Fiorentin}, {Regibo}, {Reyl{\'e}}, {Ripepi}, {Riva}, {Rixon}, {Robichon},
  {Robin}, {Roelens}, {Rohrbasser}, {Romero-G{\'o}mez}, {Rowell}, {Royer},
  {Rybicki}, {Sadowski}, {Sagrist{\`a} Sell{\'e}s}, {Sahlmann}, {Salgado},
  {Salguero}, {Samaras}, {Sanchez Gimenez}, {Sanna}, {Santove{\~n}a},
  {Sarasso}, {Schultheis}, {Sciacca}, {Segol}, {Segovia}, {S{\'e}gransan},
  {Semeux}, {Shahaf}, {Siddiqui}, {Siebert}, {Siltala}, {Slezak}, {Smart},
  {Solano}, {Solitro}, {Souami}, {Souchay}, {Spagna}, {Spoto}, {Steele},
  {Steidelm{\"u}ller}, {Stephenson}, {S{\"u}veges}, {Szabados}, {Szegedi-Elek},
  {Taris}, {Tauran}, {Taylor}, {Teixeira}, {Thuillot}, {Tonello}, {Torra},
  {Torra}, {Turon}, {Unger}, {Vaillant}, {van Dillen}, {Vanel}, {Vecchiato},
  {Viala}, {Vicente}, {Voutsinas}, {Weiler}, {Wevers}, {Wyrzykowski}, {Yoldas},
  {Yvard}, {Zhao}, {Zorec}, {Zucker}, {Zurbach}, \&
  {Zwitter}}]{2021AA...649A...1G}
{{\it Gaia} Collaboration}, {Brown}, A.~G.~A., {Vallenari}, A., {et~al.} 2021,
  \aap, 649, A1, \dodoi{10.1051/0004-6361/202039657}

\bibitem[{{Kiman} {et~al.}(2019){Kiman}, {Schmidt}, {Angus}, {Cruz}, {Faherty},
  \& {Rice}}]{2019AJ....157..231K}
{Kiman}, R., {Schmidt}, S.~J., {Angus}, R., {et~al.} 2019, \aj, 157, 231,
  \dodoi{10.3847/1538-3881/ab1753}

\bibitem[{{McMahon} {et~al.}(2021){McMahon}, {Banerji}, {Gonzalez}, {Koposov},
  {Bejar}, {Lodieu}, {Rebolo}, \& {VHS Collaboration}}]{2021yCat.2367....0M}
{McMahon}, R.~G., {Banerji}, M., {Gonzalez}, E., {et~al.} 2021, VizieR Online
  Data Catalog, II/367

\bibitem[{{Nidever} {et~al.}(2021){Nidever}, {Dey}, {Fasbender}, {Juneau},
  {Meisner}, {Wishart}, {Scott}, {Matt}, {Nikutta}, \&
  {Pucha}}]{2021AJ....161..192N}
{Nidever}, D.~L., {Dey}, A., {Fasbender}, K., {et~al.} 2021, \aj, 161, 192,
  \dodoi{10.3847/1538-3881/abd6e1}

\bibitem[{{Penoyre} {et~al.}(2020){Penoyre}, {Belokurov}, {Wyn Evans},
  {Everall}, \& {Koposov}}]{2020MNRAS.495..321P}
{Penoyre}, Z., {Belokurov}, V., {Wyn Evans}, N., {Everall}, A., \& {Koposov},
  S.~E. 2020, \mnras, 495, 321, \dodoi{10.1093/mnras/staa1148}

\bibitem[{{Stassun} {et~al.}(2019){Stassun}, {Oelkers}, {Paegert}, {Torres},
  {Pepper}, {De Lee}, {Collins}, {Latham}, {Muirhead}, {Chittidi},
  {Rojas-Ayala}, {Fleming}, {Rose}, {Tenenbaum}, {Ting}, {Kane}, {Barclay},
  {Bean}, {Brassuer}, {Charbonneau}, {Ge}, {Lissauer}, {Mann}, {McLean},
  {Mullally}, {Narita}, {Plavchan}, {Ricker}, {Sasselov}, {Seager}, {Sharma},
  {Shiao}, {Sozzetti}, {Stello}, {Vanderspek}, {Wallace}, \&
  {Winn}}]{2019AJ....158..138S}
{Stassun}, K.~G., {Oelkers}, R.~J., {Paegert}, M., {et~al.} 2019, \aj, 158,
  138, \dodoi{10.3847/1538-3881/ab3467}

\end{thebibliography}
\bibliographystyle{aasjournal}

\end{document}